\documentclass[twocolumn,showpacs,preprintnumbers,prb,amsmath,amssymb]{revtex4}

\usepackage{graphicx}
\usepackage{dcolumn}
\usepackage{bm}
\usepackage{epsfig}
\begin{document}


\title{
Spin-polarized tunneling through randomly transparent magnetic junctions:\\
Reentrant magnetoresistance approaching the Julli{\` e}re limit
}

\author{Grigory Tkachov$^1$ and Klaus Richter$^2$}
\affiliation{
$^1$Max Planck Institute for the Physics of Complex Systems, 01187 Dresden, Germany\\
$^2$Institute for Theoretical Physics, Regensburg University, 
93040 Regensburg, Germany}


\begin{abstract}
Electron conductance in planar magnetic tunnel junctions 
with long-range barrier disorder  
is studied within Glauber-eikonal approximation  
enabling exact disorder ensemble averaging 
by means of the Holtsmark-Markov method.
This allows us to address a hitherto unexplored regime 
of the tunneling magnetoresistance effect
characterized by the crossover from 
$\bm{k}_\|$-conserving to random tunneling 
($\bm{k}_\|$ is the in-plane wave vector)
as a function of the defect concentration. 
We demonstrate that such a crossover results in 
a reentrant magnetoresistance: It goes through a pronounced minimum 
before reaching disorder- and geometry-independent 
Julli{\` e}re's value at high defect concentrations.  
\end{abstract}

\maketitle

\section{Introduction}
\label{Intro}

Magnetic tunnel junctions 
with controllable relative orientation 
of the magnetization in the leads~\cite{Moodera95,Miyazaki95,Parkin04} 
are in the focus of current research motivated by 
their promising application potential as well as 
general interest in spin-dependent phenomena 
in complex condensed matter systems~\cite{Zutic04}. 
In particular, among various theoretical studies of spin-polarized transport
a large body of work has aimed at developing adequate models 
for the tunneling magnetoresistance in single-particle approximation~\cite{Slonczewski89,Bratkovsky97,MacLaren97,Mavr00,Mathon01,Butler01,Beletskii07,Heiliger07}, 
and by accounting for many-body effects due to 
electron-magnon interactions in normal~\cite{Zhang97,MacDonald98,Guinea98,Brat98,Hong02} 
and superconducting~\cite{Tkachov02, McCann02} states,   
and the influence of disorder~\cite{Tsymbal98,Tsymbal03,Mathon06,Xu06}.

The subject of the present study is 
the tunneling magnetoresistance (TMR) effect 
originating from the dependence of the tunneling current 
on the relative orientation of the magnetizations in two electrodes 
separated by a thin insulating layer~\cite{Moodera95,Miyazaki95,Parkin04}. 
Usually, the tunnel structure is designed in such a way 
that in a zero external magnetic field 
the magnetic moments are antiparallel (AP) to each other, 
and change to the parallel (P) configuration upon application 
of a weak magnetic field $B$.   
If $R(0)$ and $R(B)$ are the resistances 
in the AP and P configurations, respectively, 
the TMR ratio can be defined as $TMR\equiv (R(0)-R(B))/R(0)$, 
or as $TMR=(G^P-G^{AP})/G^P$ in terms of the corresponding conductances 
$G^{AP}=R^{-1}(0)$ and  $G^{P}=R^{-1}(B)$.
 
As the effect stems from the exchange interaction, 
it is not surprising that 
spin-dependent scattering is often seen as the main obstacle 
for achieving higher TMR ratios~\cite{Zhang97,MacDonald98,Guinea98,Brat98,Hong02,Vedyayev01}.  
Less obvious is that spin-independent elastic scattering can affect the TMR 
as well~\cite{Tsymbal98,Tsymbal03,Mathon06,Xu06}, 
in particular in the presence of structural disorder 
in the insulating barrier. 
This can be interpreted in terms of disorder-induced 
mixing of conducting channels with different wave vectors 
$\bm{k}_\|$ in the junction plane 
occuring independently for the two spin species.
On the other hand, according to Julli{\` e}re's conjecture~\cite{Julliere75} 
an increase in the amount of barrier disorder 
should eventually lead to a completely random $\bm{k}_\|$ transfer, 
with the TMR ratio depending only 
on electron spin polarizations in the magnetic leads.
The question of how the crossover between the $\bm{k}_\|$-conserving 
and Julli{\` e}re regimes actually occurs and  
which of them is most favorable for achieving a higher TMR
are the facets of a challenging problem currently under investigation.
Although some aspects of this problem have been addressed 
by numerical techniques~\cite{Tsymbal98,Tsymbal03,Mathon06,Xu06}, 
we feel that there is an apparent lack of analytical work aimed at 
proving Julli{\` e}re's conjecture from the general standpoint of 
statistical theory of quantum transport within a model-based approach.

\begin{figure}[b]
\centerline{
\epsfig{figure=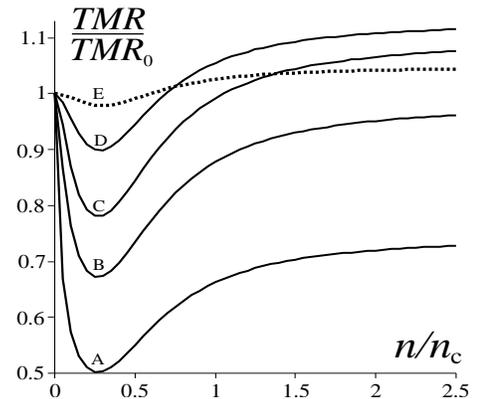, width=.7 \hsize ,height=.6 \hsize}
}
\caption{
Tunneling magnetoresistance  (TMR)
vs. defect concentration 
for different values of spin polarization: 
(A) $P=0.01$, (B) $P=0.05$, (C) $P=0.1$, (D) $P=0.2$ and 
(E) $P=0.4$. 
$TMR_0$ corresponds to a defect-free junction,
The characteristic concentration $n_c$ is defined in text (see Eq.~(\ref{n_c})).
}
\label{TMR_n}
\end{figure}

In the present work we propose an analytically solvable statistical model 
for spin-polarized tunneling describing the full disorder-driven crossover 
from the $\bm{k}_\|$-conserving to random tunneling regime.
The model assumes nonresonant long-range defects in the barrier 
and works in the thermodynamic limit where the crossover 
is controlled by the defect concentration $n$.  
The scenario of the crossover appears quite unusual 
as depicted in Fig.~\ref{TMR_n}: 
First decreasing with $n$ at low concentrations, the $TMR$  
can eventually recover that of an ideal junction, $TMR_0$, or even 
exceed it approaching an $n$-independent value for large defect concentrations. 
As we demonstrate later this value corresponds 
to Julli{\` e}re's TMR for the junctions addressed here.
To our knowledge such a reentrant TMR effect has not been 
studied previously. This finding could potentially be used for 
defect engineering of magnetic tunnel junctions.

We also find that the lower the electron spin polarization 
the stronger is the effect of the long-range disorder. 
This to some extent is in line
with the experimental observation of a rather weak TMR effect 
in  semiconductor/nonepitaxial 
iron tunnel junctions~\cite{Kreuzer02,Zenger04} 
where one may expect long-range electrostatic disorder 
in the semiconductor barrier. 
A quantitative comparison with the experimental data is 
unfortunately hindered by approximations we have to resort to 
in order to get an analytically tractable theory.

The subsequent sections give a complete account 
of our theoretical approach: 
Sec.~\ref{model} describes the model for randomly transparent barriers 
and the main approximations used.  
In Sec.~\ref{conductance} we employ the Holtsmark-Markov 
averaging procedure to calculate the spin-dependent junction conductance
and Sec.~\ref{TMR} contains the results and final discussion.   

\section{Eikonal approximation for tunneling through nonuniform barriers}
\label{model}

\begin{figure}[t]
\centerline{
\epsfig{figure=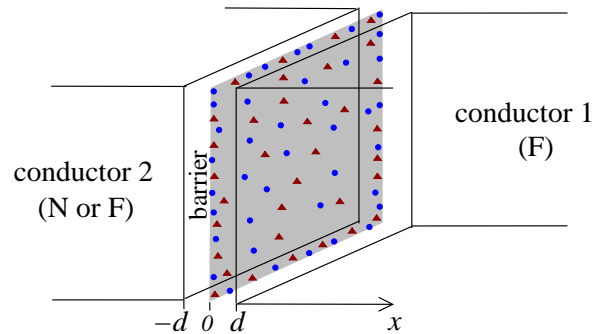, width=.9 \hsize ,height=.5 \hsize}
}
\caption{
(Color online) Lateral tunnel junction with an insulating layer of thickness $2d$ 
containing two types of defects, schematically shown as circles and triangles, 
producing a spacial barrier fluctuation described by Eq.~(\ref{U}).
System 1 is a ferromagnet, while System 2 can be either nonmagnetic or ferromagnetic.
}
\label{Geo}
\end{figure}

We consider a lateral junction between two conductors 
separated by an insulating barrier of width $2 d$ (see, Fig.~\ref{Geo}) 
modelled by a potential of the form
\begin{equation}
	{\cal U}(x,\bm{\rho})=U_0
	+\sum_{i=1}^N\left[
	U(x,|\bm{\rho}-\bm{\rho}_{i}|)-U(x,|\bm{\rho}-\bm{\rho^\prime}_i|)\right]
	\label{U}
\end{equation}
for $|x|\leq d$. 
The barrier inhomogeneity is described by the second term 
as the superposition of $N$ pairs of 
opposite-sign potentials centered at points 
$\bm{\rho}_{i}$ and $\bm{\rho^\prime}_i$ 
randomly distributed over a large junction area $A$.
We do not distinguish the $x$ coordinates 
of the defects assuming $U(x,|\bm{\rho}|)$ 
to vary smoothly with $x$ across the barrier.   
The inhomogeneous part vanishes upon averaging 
over the junction area so that it represents the lateral spatial 
fluctuation of the barrier potential around a mean value, $U_0$
(measured from the Fermi level $E_F$). 
While capturing generic features of randomly transparent barriers, 
this model, in particular, describes electrostatic disorder 
in an overall neutral insulator containing an
equal amount of donors and acceptors. 
If they are distributed homogeneously,
the Fermi level remains in the middle of the band gap~\cite{Shklovskii84} 
so that at sufficiently low bias voltage and temperature 
we can neglect resonant tunneling. 
In the situations where resonant tunneling does contribute to the TMR~\cite{Tsymbal03,Mathon05} 
our model can still be used to qualitatively study the background (nonresonant) 
contribution to the magnetoresistance as a function of the defect concentration in the barrier. 

Within linear-response theory~\cite{Kubo,Greenwood}  
we can express the junction conductance, $g_\sigma$ 
(measured in units of $e^2/h$),
for electrons with spin $\sigma=\uparrow,\downarrow$ at zero temperature as
\begin{eqnarray}
	g_\sigma=(2\pi\hbar)^2\sum\limits_{\bm{k_1}\bm{k_2}}
	|J_{\bm{k_1}\bm{k_2}}|^2\,
	\delta(E_F-E_{\sigma \bm{k_1}})\delta(E_F-E_{\sigma \bm{k_2}})  .
	\label{g}
\end{eqnarray}
The matrix elements of the current operator, $J_{\bm{k_1}\bm{k_2}}$,
evaluated in the barrier separating conductors 1 and 2 
describe spin-conserving tunneling 
between their states with wave vectors $\bm{k}_1$ and $\bm{k}_2$
at energies $E_{\sigma \bm{k_1}}$ and $E_{\sigma \bm{k_2}}$. 
The dependence of $J_{\bm{k_1}\bm{k_2}}$ on the components  $k_{jx}$
perpendicular and  $\bm{k}_{j\|}$ parallel to the interface ($j=1,2$) 
is given by~\cite{Bardeen,Harrison,Duke}
\begin{eqnarray}
  &&  
  J_{\bm{k_1}\bm{k_2}}=
  \frac{i\hbar}{2m_B}\psi_{k_{1x}}(d)\psi_{k_{2x}}(-d)
  \,W_{\bm{k_{1\|}}\bm{k_{2\|}}},
  \label{J}\\
  &&
  W_{\bm{k_{1\|}}\bm{k_{2\|}}}=A^{-1}\int 
  d\bm{\rho}\,\text{e}^{i(\bm{k_{2\|}}-\bm{k_{1\|}})\bm{\rho}}\,
  W(\bm{\rho}),
  \label{Wkk}\\
  &&
  W(\bm{\rho})=\phi_2(x,\bm{\rho})\partial_x\phi_1(x,\bm{\rho})
  -\phi_1(x,\bm{\rho})\partial_x\phi_2(x,\bm{\rho}).\,\,
  \label{W}
\end{eqnarray}
In Eq.~(\ref{J}), $m_B$ is the effective mass in the barrier. 
It is assumed that in the absence of tunneling 
the one-particle states in the leads are 
$\psi_{k_{jx}}(x)\text{e}^{i\bm{k_{j\|}}\bm{\rho} }/A^{1/2}$,
where the basis functions $\psi_{k_{jx}}(x)$
in the $x$ direction will be specified later.
The tunneling coupling is accounted for by the Wronskian, Eq.~(\ref{W}),
of the two independent solutions $\phi_{1,2}(x,\bm{\rho})$
of the Schr{\"o}dinger equation inside the barrier such 
that $\phi_j(x,\bm{\rho})$ decays into the barrier from 
side $j=1,2$~\cite{Bardeen,Harrison,Duke}.

To calculate $\phi_{1,2}(x,\bm{\rho})$ we assume that 
the inverse penetration length at the mean value of the barrier potential, 
$\kappa=(2m_BU_0/\hbar^2)^{1/2}$, satisfies the 
conditions
\begin{equation}
	\kappa\gg \sqrt{2m_B U(x,\bm{\rho})}/\hbar \quad , \quad 
	 \kappa \gg k_{j\sigma}  \; ,
	\label{high}
\end{equation}
where $k_{j\sigma}$ are the Fermi wave vectors in the leads $j=1,2$.
In this case the tunneling problem becomes effectively one-dimensional, 
which, together with the smoothness of $U(x,\bm{\rho})$ on the scale of $\kappa^{-1}$, 
allows to empoy the Glauber (eikonal) approximation 
frequently used in high-energy scattering theory~\cite{Glauber}.
Under conditions (\ref{high}) the barrier wave functions are
\begin{eqnarray}
	\phi_j(x,\bm{\rho})
        &\approx &
        \exp\biggl\{
	\kappa (\pm x-d)\pm
        \nonumber\\ 
        &
        \pm
        &
        \sum\limits_{i=1}^N
	\left[u_\pm(x,\bm{\rho}-\bm{\rho}_i)-u_\pm(x,\bm{\rho}-\bm{\rho}^\prime_i)\right] \biggr\} 
	 \, , 
	\label{phi12}\\
	u_\pm(x,\bm{\rho})
        &=&
        \frac{\kappa}{2U_0}
	\int^x_{\pm d}dx^\prime
	U(x^\prime,|\bm{\rho}|) \quad ,
	\quad
	\kappa d\gg 1 \, ,
	\label{u12}
\end{eqnarray}
where $\pm$ correspond to the functions penetrating the barrier 
from sides $j=1,2$ respectively.
Using Eqs. (\ref{phi12}) and (\ref{u12}), 
we obtain for the Wronskian (\ref{W})
\begin{eqnarray}
&&
	W(\bm{\rho})\approx W_0\, \exp\biggl\{ 
	\,\,\sum\limits_{i=1}^N[u(\bm{\rho}-\bm{\rho}^\prime_i)-u(\bm{\rho}-\bm{\rho}_i)] \biggr\} \, ,
        \label{W1}\\
&&
	u(\bm{\rho})=
        \frac{\kappa}{2U_0}
        \int\limits^d_{-d}dx 
        U(x,|\bm{\rho}|) \, .
	\label{u}
\end{eqnarray}
Here, $W_0=2\kappa\exp(-2\kappa d)$ is the Wronskian for a uniform barrier 
where the  matrix (\ref{Wkk}) is proportional to a Kronecker delta 
$\delta_{\bm{k_{1\|}}\bm{k_{2\|}}}$.
Equation (\ref{W1}) represents the main result of this section: 
In the tunneling overlap of the barrier wave functions
the lateral fluctuation of the potential 
is exponentially amplified and  
is not necessarily weak since $u$ [Eq.~(\ref{u})] 
contains the large parameter $\kappa d$. 

Equation (\ref{g}) for the conductance 
can be recast into the more convinient form
\begin{eqnarray}
	g_\sigma=\left( \frac{\pi\hbar^2}{m_B} \right)^2
	\sum_{ \bm{k_{1\|}}\bm{k_{2\|}} }
	|W_{\bm{k_{1\|}}\bm{k_{2\|}}}|^2
	\nu_{\sigma\bm{k}_{1\|}}(d)\nu_{\sigma\bm{k}_{2\|}}(-d),
	\label{g1}
\end{eqnarray}
where
\begin{equation}
\nu_{\sigma\bm{k}_{j\|}}(\pm d)=
\sum_{k_{jx}}|\psi_{k_{jx}}(\pm d)|^2\delta(E_F-E_{\sigma\bm{k}_j})
\end{equation}
are the local densities of states for given $\bm{k}_{j\|}$ and $\sigma$ at the 
boundaries of systems 1 and 2.
We now make use of the explicit expressions 
\begin{eqnarray}
\psi_{k_{1x}}(x)&=&(2/L_1)^{1/2}\sin[k_{1x}(x-d)+\gamma_1], \\
\psi_{k_{2x}}(x) & = & (2/L_2)^{1/2}\sin[k_{2x}(x+d)-\gamma_2]
\end{eqnarray}
for the basis functions in the $x$ direction,
where $L_{1,2}$ are the lengths 
of the systems. The phases $\gamma_{1,2}=\arctan(k_{1,2x}/\kappa)$ 
are determined by the boundary conditions 
$\partial_x\psi_{k_{1x}}(d)=\kappa \psi_{k_{1x}}(d)$ and 
$\partial_x\psi_{k_{2x}}(-d)=-\kappa \psi_{k_{2x}}(-d)$ 
valid for abrupt interfaces~\footnote{
Here we ignore small corrections of order of 
$U(x,|\bm{\rho}|)/U_0$.}.
Assuming furthermore spherical Fermi surfaces 
with effective mass $m$,
we find
\begin{equation}
	\nu_{\sigma\bm{k}_{j\|}}(\pm d)=\frac{2m}{\pi\hbar^2}
	\frac{(k^2_{j\sigma}-k^2_{j\|})^{1/2}\Theta(k_{j\sigma}-k_{j\|})}
	{k^2_{j\sigma}+\kappa^2-k^2_{j\|}}.
	\label{nu}
\end{equation}
To finally prepare Eq.~(\ref{g1}) for averaging over the ensemble of disorder realizations
we recast the square of the matrix elements $W_{\bm{k_{1\|}}\bm{k_{2\|}}}$ [Eq.~(\ref{Wkk})] 
as follows:
\begin{eqnarray}
&&
	|W_{\bm{k_{1\|}}\bm{k_{2\|}}}|^2=
	\int\frac{ d\bm{\delta} }{A}\,\text{e}^{i(\bm{k_{2\|}}-\bf{k_{1\|}})\bm{\delta}}
  \times
  \label{WWk}\\
&& \times
  \int\frac{ d\bm{\rho} }{A}\, 
  W\left( 
  \bm{\rho}+\frac{ \bm{\delta} }{ 2 }
  \right)
  W\left( 
  \bm{\rho}-\frac{ \bm{\delta} }{ 2 }
  \right)\to
  \nonumber\\
  &&
	\to\int\frac{ d\bm{\delta} }{A}\,
  \text{e}^{i(\bm{k_{2\|}}-\bf{k_{1\|}})\bm{\delta}}
  \left\langle 
  W\left( 
  \bm{\rho}+\frac{ \bm{\delta} }{ 2 }
  \right)
  W\left( 
  \bm{\rho}-\frac{ \bm{\delta} }{ 2 }
  \right)
  \right\rangle_{conf} \; .
  \nonumber
\end{eqnarray}
Here averaging over different points $\bm{\rho}$ on area $A$ 
is replaced by configurational averaging over uniformly distributed
defect positions $\bm{\rho}_{i}$ and $\bm{\rho^\prime}_i$.   

\section{Configurational averaging and spin-dependent conductance}
\label{conductance}

\subsection{Averaging procedure}

To evaluate the correlation function of the Wronskian in Eq.~(\ref{WWk}) 
for large $N$ and $A$ we employ the Holtsmark-Markov averaging procedure~\cite{Chandra} 
implemented as follows:
\begin{widetext}
\begin{eqnarray}
	\langle W(\bm{\rho}_1)W(\bm{\rho}_2)\rangle & = & W^2_0
	\left[\int\frac{d\bm{\rho}^\prime}{A}\,
	\text{e}^{u(\bm{\rho}_1-\bm{\rho}^\prime)+u(\bm{\rho}_2-\bm{\rho}^\prime)} 
	\right]^N
	\left[\int\frac{d\bm{\rho}}{A}\,
	\text{e}^{-u(\bm{\rho}_1-\bm{\rho})-u(\bm{\rho}_2-\bm{\rho})} 
	\right]^N
	\nonumber\\
	&=&
	W^2_0
	\left\{
	1-\frac{n}{N}\int d\bm{\rho}^\prime
	\left[
	1-\text{e}^{u(\bm{\rho}_1-\bm{\rho}^\prime)+u(\bm{\rho}_2-\bm{\rho}^\prime)}
	\right]
	\right\}^N
	\left\{
	1-\frac{n}{N}\int d\bm{\rho}
	\left[
	1-\text{e}^{-u(\bm{\rho}_1-\bm{\rho})-u(\bm{\rho}_2-\bm{\rho})}
	\right]
	\right\}^N
	\nonumber\\
       &\approx & W^2_0
	\exp\left\{
	-n\int d\bm{\rho}^\prime
	\left[
	1-\text{e}^{u(\bm{\rho}_1-\bm{\rho}^\prime)+u(\bm{\rho}_2-\bm{\rho}^\prime)}
	\right]
	\right\}
	\exp\left\{
	-n\int d\bm{\rho}
	\left[
	1-\text{e}^{-u(\bm{\rho}_1-\bm{\rho})-u(\bm{\rho}_2-\bm{\rho})}
	\right]
	\right\} \; .
	\label{WW}
\end{eqnarray}
\end{widetext}
In the last step we took the limit $N,A\to\infty$, 
introducing a finite defect concentration $n=N/A$. 
Clearly, the above correlation function becomes 
independent of the "center-of-mass" position 
$(\bm{\rho}_1+\bm{\rho}_2)/2$ and can be rewritten as
\begin{eqnarray}
	\langle W(\bm{\rho}_1)W(\bm{\rho}_2)\rangle=\langle W^2\rangle
	\exp{\{ n[ C(\bm{\rho}_1-\bm{\rho}_2) - C(0)]\}}\,\,
\label{WW1}
\end{eqnarray}
with
\begin{eqnarray}
&
\langle W^2\rangle=W^2_0 
\exp\biggl\{
8\pi n\int \rho d\rho\sinh^2 u(\rho) 
\biggr\} \, ,
& 
\label{W_sq}\\
&
C(\bm{\delta})=2\int d\bm{\rho}\,
\cosh\left[
u\left(\bm{\rho}+\bm{\delta}/2\right)+
u\left(\bm{\rho}-\bm{\delta}/2\right)
\right] \, ,
&
\label{C} 
\end{eqnarray}
where $ \bm{\delta}=\bm{\rho}_1-\bm{\rho}_2$.
For the matrix elements (\ref{WWk}) we then obtain
\begin{equation}
|W_{\bm{k_{1\|}}\bm{k_{2\|}}}|^2=\langle W^2 \rangle
\int\frac{ d\bm{\delta} }{A}\,
\text{e}^{-n[C(0)-C(\bm{\delta})]-i(\bm{k_{1\|}}-\bf{k_{2\|}})\bm{\delta}} \, .
\nonumber
\end{equation}
Since the integration involves a rapidly oscillating function, 
we expand $C( \bm{\delta} )$ in powers of $\bm{\delta}$,
\begin{equation}
C( \bm{\delta} )\approx C(0)-
\bm{\delta}^2 \int 2\pi\rho d\rho\,
(du/d\rho)^2\cosh 2u(\rho) \, . 
\end{equation}
Then the integration can be easily performed yielding
\begin{eqnarray}
	|W_{\bm{k_{1\|}}\bm{k_{2\|}}}|^2\approx\langle W^2 \rangle
	\frac{2\pi\rho_c^2}{A}\,
	\text{e}^{-\frac{1}{2}(\bm{k_{1\|}}-\bm{k_{2\|}})^2\rho^2_c} \, ,
	\label{WWk1}
\end{eqnarray}
where the radius
\begin{equation}
	\rho_c=
	\left[
	4\pi n \int\rho d\rho\,(du/d\rho)^2\cosh 2u(\rho)
	\right]^{-1/2}
	\label{r}
\end{equation}
characterizes the spatial decay of the correlations: 
\begin{equation}
\langle W(\bm{\rho}_1)W(\bm{\rho}_2)\rangle
\approx\langle W^2 \rangle\exp[-(\bm{\rho}_1-\bm{\rho}_2)^2/2\rho^2_c] \, .
\end{equation}
Equation (\ref{g1}) for the conductance then finally reads
\begin{eqnarray}
	&&
	g_\sigma=
	\left(\frac{\hbar^2}{2m_B}\right)^2\frac{\langle W^2\rangle A\rho_c^2}{2\pi}\times
	\label{general}\\
	&&
	\times
	\int d\bm{k}_{1\|}d\bm{k}_{2\|}
	\text{e}^{-\frac{1}{2}(\bm{k_{1\|}}-\bm{k_{2\|}})^2\rho_c^2}
	\,\,\nu_{\sigma\bm{k}_{1\|}}(d)\nu_{\sigma\bm{k}_{2\|}}(-d) \; .
	\nonumber
\end{eqnarray}
The effect of the disorder depends on the dimensionless parameters 
$\rho_ck_{1\sigma}$  and $\rho_ck_{2\sigma}$ 
controlling the crossover between the $\bm{k}_\|$-conserving 
and random tunneling regimes.

\subsection{$\bm{k}_\|$-conserving tunneling}

For weak disorder ($\rho_ck_{1,2\sigma}\gg 1$) 
the matrix elements in Eq.~(\ref{general}) 
have a sharp maximum at $\bm{k}_{1\|}=\bm{k}_{2\|}$ 
and hence can be integrated out. 
This yields the Landauer-type formula 
\begin{eqnarray}
	g_\sigma & = & \frac{A}{(2\pi)^2}\int d\bm{k}_{\|}T_\sigma(\bm{k}_{\|}),
	\label{g_c}\\
T_\sigma(\bm{k}_{\|}) & = & \left(  \frac{ \pi W_0\hbar^2}{m} \right)^2
\nu_{\sigma\bm{k}_\|}(d)\nu_{\sigma\bm{k}_\|}(-d) \, ,
\end{eqnarray}
with $W_0$ from Eq.~(\ref{W1}) and $T_\sigma(\bm{k}_{\|})$ being 
the transmission probability for a uniform rectangular barrier.

\subsection{Random-momentum tunneling}

This regime is reached in the limit $\rho_ck_{1,2\sigma}\ll 1$
when the matrix elements become momentum 
independent. Then the integrations over  $\bm{k}_{1\|}$ and $\bm{k}_{2\|}$ 
can be done separately, 
yielding the conductance as the product of the local densities of states (DOS),
$\nu_{\sigma}(\pm d)=\int d\bm{k}_{\|}\nu_{\sigma\bm{k}_\|}(\pm d)/(2\pi)^2$:
\begin{eqnarray}
        &&
	g_\sigma=(2\pi)^3\left(\frac{\hbar^2}{2m_B}\right)^2\langle W^2\rangle A\rho_c^2
	\,\,\nu_{\sigma}(d)\nu_{\sigma}(-d) \, ,
	\label{g_r}\\
        &&
	\nu_{\sigma}(\pm d)=\nu^{(\text{bulk})}_{j\sigma}
	\left[1-(\kappa/k_{j\sigma})\arctan (k_{j\sigma}/\kappa)
	\right] \, .
	\label{LDOS}
\end{eqnarray}
The presence of the local DOS in Eq.~(\ref{g_r}) 
reflects the sharpness of the interfaces at $x=\pm d$. 
The difference between the local DOS  [Eq.~(\ref{LDOS})] 
and the corresponding DOS in the bulk, $\nu^{(\text{bulk})}_{j\sigma}=mk_{j\sigma}/\pi^2\hbar^2$
is particularly pronounced for a high 
barrier with $k_{j\sigma}/\kappa\ll 1$:
\begin{equation}
\nu_{\sigma}(\pm d)\approx
\nu^{(\text{bulk})}_{j\sigma}
\left( \frac{ k_{j\sigma} }{\sqrt{3}\kappa} \right)^2\ll \nu^{(\text{bulk})}_{j\sigma}.
\label{LDOS1}
\end{equation}
In the next section, we demonstrate that the random-momentum tunneling results in 
Julli{\` e}re's magnetoresistance.    

\section{Tunneling spin polarization and magnetoresistance}
\label{TMR}

It is instructive to consider first the effect of the barrier disorder 
on the spin polarization of the tunneling current 
between a ferromagnet and a nonmagnetic conductor.
For small bias voltages the current spin polarization 
can be expressed in terms of 
the spin-resolved conductances:   
\begin{equation}
P_J= (g_\uparrow -g_\downarrow)/
(g_\uparrow +g_\downarrow) \, .
\label{eq:Pj}
\end{equation}
We assume that system 1 is a Stoner ferromagnet 
whose electron spin polarization $P$ is characterized 
by the bulk DOS for spin-up (majority) and spin-down (minority) carriers,
\begin{equation}
P=(\nu^{(\text{bulk})}_{1\uparrow}-\nu^{(\text{bulk})}_{1\downarrow})/
(\nu^{(\text{bulk})}_{1\uparrow}+\nu^{(\text{bulk})}_{1\downarrow}) \, .
\label{Pnu}
\end{equation}
The Fermi wave vectors are parametrized as
\begin{eqnarray}
&
k_{1\uparrow}=k\sqrt{1+\Delta_P},\quad 
k_{1\downarrow}=k\sqrt{1-\Delta_P},
& 
\label{k_F}\\
&
\Delta_P=2P/(1+P^2),
&
\label{Delta}
\end{eqnarray}  
where $\Delta_P$ is the dimensionless band spin-splitting,
and $k=\sqrt{(k^2_{1\uparrow}+k^2_{1\downarrow})/2}$. 
System 2 has a spin-independent DOS, 
$\nu^{(\text{bulk})}_{2\sigma}=\nu^{(\text{bulk})}_{2}$,
and Fermi wave vectors $k_{2\sigma}=k$.

\begin{figure}[t]
\centerline{
\epsfig{figure=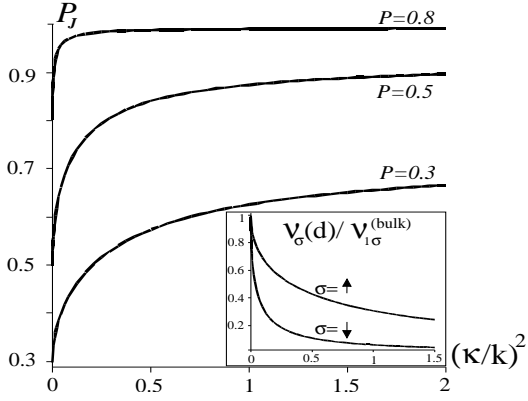, width=.8 \hsize ,height=.6 \hsize}
}
\caption{Current spin polarization, Eq.~(\ref{eq:Pj}), as a function of
 $(\kappa/k)^2$ characterizing the
strength of the mean barrier potential $U_0$ for different values of
the electron polarization, Eq.~(\ref{Pnu}). 
Inset: Spin-resolved local DOS at the boundary of the spin injector versus
$(\kappa/k)^2$ for $P=0.3$; $\kappa$ and $k$ are defined in text.}
\label{PJ}
\end{figure}

In the random tunneling regime, Eq.~(\ref{g_r}), the current spin polarization, 
$
P_J=(\nu_{\uparrow}(d)-\nu_{\downarrow}(d))/(\nu_{\uparrow}(d)+\nu_{\downarrow}(d)),
$
reflects the electron spin polarization at the surface of the spin injector,  
As shown in Fig.~\ref{PJ}, $P_J$ increases 
with the ratio $(\kappa/k)^2$ characterizing the
strength $U_0$ of the mean barrier potential. 
Such an enhancement can be traced back to 
the behavior of the spin-resolved local DOS 
(see inset in Fig.~\ref{PJ}). 
Although both, $\nu_{\uparrow}(d)$ and $\nu_{\downarrow}(d)$
are suppressed compared to the bulk values, 
the difference between them increases
leading to the higher surface spin polarization. 

\begin{figure}[t]
\centerline{
\epsfig{figure=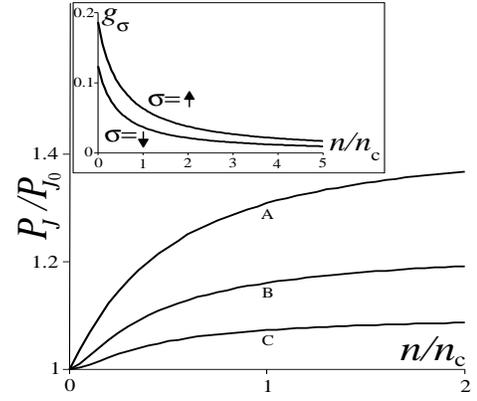, width=.7 \hsize ,height=.6 \hsize}
}
\caption{
Current spin polarization vs. defect concentration in units of 
$P_{J_0}=P_J(n=0)$ and $n_c=n(\rho_c k)^{2}$: 
(A) $P=0.1$, (B) $P=0.3$, and (C) $P=0.45$. 
Inset: spin-dependent conductances for $P=0.1$;
For convenience, they are normalized to spin-independent factor 
$(m/m_B)^2\langle W^2\rangle Ak^2/\kappa^4$.
}
\label{PJ_n}
\end{figure}

In what follows we address exclusively 
systems with large values of $(\kappa/k)^2$. 
Then the double integral in Eq.~(\ref{general}) 
can be transformed to a single one 
in position representation~\footnote{
This is a fairly good approximation for $(\kappa/k)^2\geq 5$.}: 
\begin{eqnarray}
	&&
        g_\sigma=\frac{m^2}{m^2_B}\frac{(k_{1\sigma}k_{2\sigma})^{3/2}}{\kappa^4} \langle W^2\rangle A\times 
	\label{g_strong} \\
        &&
	\times\int_0^\infty d\delta\, \exp\left(-\frac{\delta^2}{2\rho^2_c} \right)\,
	\frac{J_{3/2}(k_{1\sigma}\delta)J_{3/2}(k_{2\sigma}\delta)}{\delta^2} \, , 
        \nonumber
\end{eqnarray}
where $J_{3/2}(x)$ is a Bessel function. 
It follows from Eq.~(\ref{g_strong}) 
that $P_J$ exceeds the clean-barrier value $P_{J_0}=P_J(n=0)$ 
at any finite defect concentration $n$ (see Fig.~\ref{PJ_n}). 
There is a characteristic value $n_c$ 
related to the correlation radius, Eq.~(\ref{r}) as 
\begin{equation}
n_c=n(\rho_ck)^{2} \, .
\label{n_c}
\end{equation} 
The inset in Fig.~\ref{PJ_n} shows the corresponding behavior 
of the majority and minority electron conductances. 

We wish to understand to what extent the behavior 
of the tunneling spin polarization, $P_J$, correlates with 
that of the tunneling magnetoresistance. To this end
we consider a junction between 
two identical Stoner ferromagnets for which the TMR ratio is
$TMR=1-\sum_\sigma g^{AP}_\sigma/
\sum_\sigma g^P_\sigma$, 
where $g^P_\sigma$ and $g^{AP}_\sigma$ are both given by Eq.~(\ref{g_strong}) 
with $k_{1\sigma}$ and $k_{2\sigma}$ defined in the following way. 
In the P case, we choose the Fermi wave vectors of 
the majority and minority electrons to be 
$k_{maj}=k_{1\uparrow}=k_{2\uparrow}$ and $k_{min}=k_{1\downarrow}=k_{2\downarrow}$, 
while for AP $k_{maj}=k_{1\uparrow}=k_{2\downarrow}$ and
$k_{min}=k_{1\downarrow}=k_{2\uparrow}$. 
We again use the parametrization  
$k_{maj}=k\sqrt{1+\Delta_P}$ and $k_{min}=k\sqrt{1-\Delta_P}$ with 
$k$ and $\Delta_P$ defined earlier in Eqs.~(\ref{k_F}) and (\ref{Delta}).

The dependence of the TMR on the defect concentration 
was already discussed in the introduction (see, Fig.~\ref{TMR_n}).
Unlike $P_J(n)$ it is non-monotonic with a minimum at $n\approx 0.25n_c$ 
most pronounced for relatively low electron polarization (curves A-D).  
However, similar to $P_J(n)$ the TMR increases for $n/n_c>1$ 
saturating at the value 
\begin{equation}
	TMR=(\nu_{maj}-\nu_{min})^2/(\nu^2_{maj}+\nu^2_{min})
	\label{Julliere}
\end{equation}
depending solely on the local DOS of the majority
and minority electrons, $\nu_{maj}$ and $\nu_{min}$. 
Equation (\ref{Julliere}) follows from the asymptotic 
expression (\ref{g_r}) and definitions of $k_{maj}$ and $k_{min}$, 
reproducing the Julli{\` e}re's result for a sharp barrier. 

\begin{figure}[b]
\centerline{
\epsfig{figure=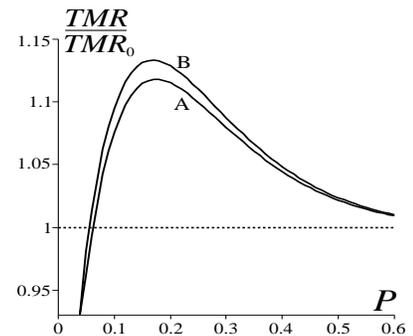, width=.6 \hsize ,height=.5 \hsize}
}
\caption{Tunneling magnetoresistance as a function of bulk spin polarization 
for (A) $n=2.5n_c$ and (B) $n=25n_c$. 
}
\label{TMR_P}
\end{figure}

Another surprising feature of the TMR is that 
for strongly disordered barriers with $n/n_c>1$ 
the TMR ratio turns out to be a non-monotonic function of  
electron spin polarization $P$ as shown in Fig.~\ref{TMR_P}. 
By choosing the ferromagnets with $P\approx 0.15-0.2$
one can take advantage of the barrier disorder 
to enhance the TMR response.

In conclusion, we discuss the applicability of our results. 
They are valid for the homogeneous defect distribution 
in a large-area barrier so that both the current spin polarization 
and TMR are independent of the junction cross-sectional geometry. 
They also do not depend on a concrete form of the defect potential 
$U(x,\bm{\rho})$ provided the eikonal approximation holds 
(see Sec.~\ref{model}). 
Let us estimate the correlation radius $\rho_c$ [Eq.~(\ref{r})] and 
the characteristic concentration $n_c$ [Eq.~(\ref{n_c})]
for a system of compensating donors and acceptors with charges $\pm e$ 
distributed on a plane $x=0$ in the middle of the barrier 
(so-called $\delta$ doping~\cite{Jansen00}). 
Since the defects release no charge carriers, 
we consider the screening of the defect potential $U(x,\bm{\rho})$ 
by the conducting electrodes. 
To simplify the calculations we assume 
ideally screening electrodes described by 
the boundary conditions $U(x=\pm d,\bm{\rho})=0$ 
for the Poisson equation 
$\triangle U=(4\pi e^2/\epsilon)\delta(x)\delta(\bm{\rho})$
where $\epsilon$ is the dielectric constant of the barrier material. 
The solution of such a boundary problem is a superposition of 
the Coulomb potentials of the defect charge and the image ones, 
\begin{equation}
	U(x,\bm{\rho})=\frac{e^2}{\epsilon}\sum_{n=-\infty}^{\infty}
	\frac{(-1)^n}{ \sqrt{ (x-2dn)^2 + \bm{\rho}^2 } },
	\label{mirror}
\end{equation}
where terms with $n=\pm 1,\pm 2,...$ represent
an infinite sign-alternating series of image charges
at points $x=2dn$ generated by multiple "reflections" 
in two "mirrors" $x=\pm d$. 
The boundary conditions are satisfied due to 
the even and odd terms cancelling each other at $x=\pm d$.  
For the estimate it suffice to know the asymptotic formula 
for $\rho\gg d$, obtained by replacing 
$\sum_n\to\int dn$ and $(-1)^n\to \cos(\pi n)$ in Eq.~(\ref{mirror}). 
The integration yields 
\begin{equation}
	U(x,\bm{\rho})\approx\frac{e^2}{d\epsilon}\cos\left(\frac{\pi x}{2d}\right)
	K_0\left( \frac{\pi |\bm{\rho}|}{2d} \right),
	\label{mirror1}
\end{equation}
where $K_0(x)$ is the modified Bessel function of second kind.
 
It can be checked numerically that Eq.~(\ref{mirror1}) 
is also a fairly good approximation for $\rho\leq d$, 
except for the immediate vicinity of the charge location $x=0,\rho=0$.  
Since for our system $\kappa d\gg 1$ [see, Eq.~(\ref{u12})],  
the $x$ dependence of the potential, Eq.~(\ref{mirror1}) 
is smooth on the scale of the electronic penetration length $\kappa^{-1}$. 
This justifies our quasiclassical approach, in particular, 
allowing us to neglect smearing of the defect distribution about the plane $x=0$ 
as long as they are still deep inside the barrier.  
For experimental situations where charged defects 
occur at the barrier boundaries or inside the conductors, 
our model needs to be modified to account for the Thomas-Fermi screening 
and scattering from such defects.

Using Eqs.~(\ref{r}) and (\ref{mirror1}) we find the correlation radius,
\begin{equation}
	\rho_c=\frac{\kappa a_B}
	{8 
	\left[\pi n	\int^\infty_0 \xi d\xi \left( \frac{dK_0(\xi)}{d\xi} \right)^2
	\cosh\left(
	\frac{8K_0(\xi)}{\pi \kappa a_B}
	\right) \right]^{1/2}
	}, 
	\label{r_est}
\end{equation}
where $a_B=\epsilon\hbar^2/e^2m_B$ is the effective Bohr radius.  
It must satisfy the condition of a weak potential fluctuation 
$\kappa a_B\gg 1$, equivalent to the first inequality in 
Eq.~(\ref{high}). 
For instance, if we take  $\kappa a_B=8$, 
then $\rho_c\approx n^{-1/2}$ up to a numerical factor,
i.e. the correlation radius is roughly 
the average distance between the defects.   
Consequently, the characteristic defect concentration 
$n_c$ at which the TMR approaches Julli{\` e}re's limit 
is related to the Fermi wave vector $k$ as 
$n_c\approx k^2$. 
This condition is  easier to meet  for 
magnetic semiconductors  than for ferromagnetic transition metals. 
As we saw, however, the long-range barrier disorder can affect 
the TMR at significantly lower concentrations.   

We thank J. Fabian, D. Ryndyk, D. Weiss and M. Wimmer  
for stimulating discussions.
The work was supported by the Deutsche Forschungsgemeinschaft within
SFB 689.
%



\begin{thebibliography}{00}

\bibitem{Moodera95}
J. S. Moodera, L. R. Kinder, T. M. Wong, and R. Meservey, Phys. Rev. Lett. {\bf 74}, 3273 (1995).

\bibitem{Miyazaki95}
T. Miyazaki and N. Tezuka, J. Magn. Magn. Mater. {\bf 139}, L231 (1995).

\bibitem{Parkin04}
S. Parkin, C. Kaiser, A. Panchula, P. Rice, B. Hughes, M. Samant, and S.-H. Yang,
{\it Nature Mat.} {\bf 3}, 862 (2004).

\bibitem{Zutic04}
I. \u{Z}uti\'{c}, J. Fabian, and S. Das Sarma, Rev. Mod. Phys. {\bf 76}, 323 (2004).

\bibitem{Slonczewski89}
J.C. Slonczewski, Phys. Rev. B {\bf 39}, 6995 (1989).

\bibitem{Bratkovsky97}
A. M. Bratkovsky, Phys. Rev. B {\bf 56}, 2344 (1997).
 
\bibitem{MacLaren97}
J. M. MacLaren, X.-G. Zhang, and W. H. Butler, Phys. Rev. B {\bf 56}, 11827 (1997).

\bibitem{Mavr00}
P. Mavropoulos, N. Papanikolaou, and P. Dederichs, Phys. Rev. Lett. {\bf 85}, 1088 (2000).

\bibitem{Mathon01}
J. Mathon and A. Umerski, Phys. Rev. B {\bf 63}, 220403 (2001).

\bibitem{Butler01}
W. Butler, X.-G. Zhang, T. Schulthess, and J. MacLaren, Phys. Rev. B {\bf 63}, 
054416 (2001).

\bibitem{Beletskii07}
N. N. Beletskii, G. P. Berman, A. R. Bishop, S. A. Borysenko, and V. M. Yakovenko, 
Phys. Rev. B {\bf 75}, 174418 (2007). 

\bibitem{Heiliger07}
C. Heiliger, M. Gradhand, P. Zahn, and I. Mertig,
Phys. Rev. Lett. {\bf 99}, 066804 (2007). 

\bibitem{Zhang97}
S. Zhang, P.M. Levy, A.C. Marley, and S.S.P. Parkin, Phys. Rev. Lett. {\bf 79}, 3744 (1997).


\bibitem{MacDonald98}
A. H. MacDonald, T. Jungwirth, and M. Kasner
Phys. Rev. Lett. {\bf 81}, 705 (1998).

\bibitem{Guinea98}
F. Guinea, Phys. Rev. B {\bf 58}, 9212 (1998).

\bibitem{Brat98}
A.M.~Bratkovsky, Appl. Phys. Lett. {\bf 72}, 2334 (1998).

\bibitem{Hong02}
J. Hong, R. Q. Wu, and D. L. Mills, 
Phys. Rev. B {\bf 66}, 100406 (2002).


\bibitem{Tkachov02}
G. Tkachov, E. McCann, and V.I. Fal'ko, Phys. Rev. B {\bf 65},  024519 (2002); 
also in "Recent Progress in Many-Body Theories" edited by R.F. Bishop, T. Brands, K.A. Gernoth, N.R. Walet and Y. Xian, {\bf 6}, p. 60 (World Scientific, 2002).

\bibitem{McCann02}
E. McCann, G. Tkachov, and V.I. Fal'ko, Physica E {\bf 12}, 938 (2002).


\bibitem{Tsymbal98}
E. Yu. Tsymbal and D. G. Pettifor, Phys. Rev. B {\bf 58}, 432 (1998).

\bibitem{Tsymbal03}
E. Y. Tsymbal, O. Mryasov, and P. R. LeClair, J. Phys.: Condens. Matter {\bf 15}, R109 (2003). 

\bibitem{Mathon06}
J. Mathon and A. Umerski, Phys. Rev. B {\bf 74}, 140404 (2006).

\bibitem{Xu06}
P. X. Xu, V. M. Karpan, K. Xia, M. Zwierzycki, I. Marushchenko, and P. J. Kelly, 
Phys. Rev. B {\bf 73}, 180402(R) (2006).

\bibitem{Vedyayev01}
A. Vedyayev, D. Bagrets, A. Bagrets, and B. Dieny, Phys. Rev. B {\bf 63}, 064429 (2001).

\bibitem{Julliere75}
M. Julli{\` e}re, Phys. Lett. {\bf 54}A, 225 (1975).

\bibitem{Kreuzer02}
S. Kreuzer, J. Moser, W. Wegscheider, D. Weiss, M. Bichler, and D. Schuh,
Appl. Phys. Lett. {\bf 80}, 4582 (2002). 


\bibitem{Zenger04}
M. Zenger, J. Moser, W. Wegscheider, D. Weiss, and T. Dietl, 
J. Appl. Phys. {\bf 96}, 2400 (2004).

\bibitem{Shklovskii84}
See, e.g. B. I. Shklovskii and A. L. Efros, 
{\em Electronic Properties of Doped Semiconductors}, (Springer, Berlin, 1984). 

\bibitem{Mathon05}
J. Mathon and A. Umerski, Phys. Rev. B {\bf 71}, 220402 (2005).

\bibitem{Kubo}
R. Kubo, J. Phys. Soc. Japan {\bf 12}, 570 (1957).

\bibitem{Greenwood}
D. A. Greenwood, Proc. Phys. Soc. London {\bf 71}, 585 (1958).

\bibitem{Bardeen}
J. Bardeen, Phys. Rev. Lett. {\bf 6}, 57 (1961).

\bibitem{Harrison}
W. A. Harrison, Phys. Rev. {\bf 123}, 85 (1961).

\bibitem{Duke}
C. B. Duke,{\it Tunneling in Solids, Solid State Physics Suppl.} 
(Academic, New York, 1969).

\bibitem{Glauber}
R. J. Glauber, Phys. Rev. {\bf 100}, 242 (1955).

\bibitem{Chandra}
See, e.g. S.~Chandrasekhar, Rev.~Mod.~Phys. {\bf 15}, 1 (1943).

\bibitem{Jansen00}
R. Jansen and J. S. Moodera, Phys. Rev. B {\bf 61}, 9047 (2000).






\end{thebibliography}
\end{document}